\documentclass[pre,twocolumn,showpacs,superscriptaddress] {revtex4}
\usepackage{amsmath,amssymb}
\usepackage{graphics,graphicx}
\usepackage{dcolumn,bm}
\usepackage{psfrag}

\newcommand{\udg}
{\affiliation{Departamento de F\'isica, Universidad de Guadalajara,
Guadalajara, Jalisco, Mexico}}

\begin{document}
  
 \title
 {
Zero temperature ordering dynamics in two dimensional BNNNI model}

 \author
 {Soham Biswas}
 \email{soham.biswas@academicos.udg.mx}
\udg
 \author {Mauricio Martin Saavedra Contreras }

\udg

\begin{abstract}
We investigate  the  dynamics of a two dimensional bi-axial next nearest neighbour
Ising  (BNNNI) model following
a quench to zero temperature. The Hamiltonian is given by
$H = -J_0\sum_{i,j=1}^L [(S_{i,j}S_{i+1,j}+S_{i,j}S_{i,j+1}) -\kappa (S_{i,j}S_{i+2,j} + S_{i,j}S_{i,j+2})]$ . 
 For $\kappa <1$, the system does not reach the equilibrium ground state and keep evolving in active states for ever. 
 For $\kappa \geq 1$, though the system reaches a final state, but it do not reach the ground state always and freezes to 
 a striped state with a finite probability like two dimensional ferromagnetic Ising model and ANNNI model. The overall dynamical behaviour for $\kappa > 1$ and $\kappa =1$ is quite different. 
  The residual energy decays in a power law for both $ \kappa >1$ and $\kappa =1$ from which the dynamical exponent $z$ have been estimated. 
 The persistence probability shows algebraic decay for $\kappa > 1$ with an exponent $\theta = 0.22 \pm 0.002$ while the
 dynamical exponent for ordering $z=2.33\pm 0.01$. For $\kappa =1$, the system belongs to a completely different dynamical class with $\theta = 0.332 \pm 0.002$ and $z=2.47\pm 0.04$. 
 We have computed the freezing probability for different values of $\kappa$. We have also studied the 
 decay of autocorrelation function with time for different regime of $\kappa$ values. 
 The results have been compared with that of the two dimensional ANNNI model. 

\vskip 0.5cm

\end{abstract}

\maketitle
\section{Introduction}

The non-equilibrium process is very complex and critical to understand \cite{privman,soths}.
In case of non-equilibrium dynamics the probability distributions are not simply the Boltzmann distributions (as
in case of equilibrium process) and change at each and every time step.  A considerable 
interest has been developed for studying the dynamics of Ising spin systems
and has emerged as a rich field of present-day research. 
 When a system is close to the critical point, anomalies can occur in a large variety of dynamical 
properties and models having identical static critical behavior may display different
 behavior when dynamic critical phenomena are considered \cite{Ho_Ha}.
 The fate and the behaviour of Ising spin system following a deep
quench below the critical temperature has been one of the central topic 
of interest in the study of the non-equilibrium dynamics for last two decades. 
In a quenching process, the system has a disordered initial configuration
 which corresponds to a very high temperature ($T \rightarrow \infty$) and its temperature is suddenly dropped.
Systems quenched from a disordered phase into an ordered phase do not order instantaneously. Instead, the
length scale of ordered regions grows with time as the different broken symmetry phases compete each other to select the equilibrium state \cite{bray}.

One dimensional Ising  model with nearest neighbour interaction does not have any non-trivial phase transition and a zero temperature quench  of the  Ising model
ultimately leads to the equilibrium configuration, i.e., all spins point up (or down). 
The system coarsen evolving according to the usual Glauber dynamics resulting quite a few interesting phenomena like domain growth \cite{gunton,bray},
 persistence \cite{satya1,derrida,stauffer,krap1,Krap_Redner} etc. 
 The average domain size $D$ increases in time $t$ as $D(t)\sim t^{1/z}$, where $z$ is the dynamical exponent associated with the growth.
In two or higher dimensions, however, following a quench to zero temperature the system does not always reach the ground state \cite{Krap_Redner} 
though these scaling relations still hold good.
Not only the domain growth  phenomenon, but another important dynamical behavior which have been commonly studied is persistence, 
the tendency of a spin in an Ising system to remain in its original state following a quench to
zero temperature \cite{satya1,derrida}. In Ising model, in a zero temperature quench, persistence is measured by the probability that a spin has not flipped till time $t$ and
shows a power law behavior, i.e $P(t)\sim t^{-\theta}$. $\theta$ is called the persistence exponent and is unrelated to any other known static or 
dynamic exponents \cite{satya1,stauffer}.

Dynamical behaviour of Ising spin systems change drastically in the presence of competing interactions. To study the
effect of the competing interaction on the dynamical behaviour, the simple Ising model with a competing next nearest
neighbor interaction has been studied earlier in both one and two dimensions \cite{redner,sdg_ps,barma,annni}.
Competing interactions could also be present in the system if 
spins have random long range interactions which are quenched in nature \cite{boyer,bsnet}.

In one dimensional, the simplest example of Ising spin system with competing interaction is the ANNNI (Axial next
nearest neighbour Ising) model with L spins which can be described by the Hamiltonian
\[
H = -J\sum_{i=1}^L (S_iS_{i+1} - \kappa S_i S_{i+2}).
\]
In this model it  was found that for $\kappa < 1$,  under a zero temperature quench with
single spin flip Glauber dynamics, the system does not reach  its true ground state. (The ground state is ferromagnetic
for $\kappa < 0.5$, antiphase for $\kappa> 0.5$, and highly degenerate at $\kappa=0.5$ \cite{selke}). 
On the other hand, after some initial short time, the number of domain walls do not decay but they remain mobile 
at all times. That makes the persistence probability go to
zero in a stretched exponential manner. However for $\kappa > 1$, although the
system reaches the ground state after long time, the dynamical exponent and the persistence exponent are both different from 
that of the Ising model with nearest neighbour interactions only \cite{sdg_ps}.

In two or higher dimensions, zero temperature quenching dynamics of Ising model only with the nearest neighbour interaction is also interesting in its own merit. 
The system does not reach the ground state always and frozen-in striped states appear \cite{Krap_Redner}. 
In three dimension, the system never reaches the ground state \cite{3dising}. 
Different interesting dynamical behaviors inspired the study of zero temperature Glauber dynamics for the two dimensional ANNNI (Axially Next Nearest neighbour Ising) 
model in which competing interaction is present along one (horizontal) direction.  
The Hamiltonian of the model on $L\times L $ lattice is given by
\begin{equation}
H = -J_0\sum_{i,j=1}^L S_{i,j}S_{i+1,j} - J_1\sum_{i,j=1} [S_{i,j} S_{i,j+1} -\kappa S_{i,j} S_{i,j+2}].
\label{annni2d}
 \end{equation}
 For $\kappa <1$, this system does not reach the equilibrium ground state (The ground state is ferromagnetic for $\kappa< 0.5$ and for $\kappa > 0.5$ antiphase order exist 
 only in the horizontal direction. On the other hand, the vertical direction is always ferromagnetic) but slowly evolves to a metastable state.
 For $\kappa \geq 1$, both the persistence probability and the number of domain walls show algebraic decay. 
For $\kappa > 1$, the system shows a behaviour similar to the 
two dimensional ferromagnetic Ising model in the sense that it freezes to 
a striped state with a finite probability. 
However for $\kappa =1$, the system belongs to a completely different
dynamical class and it always evolves to the true ground state with  the
persistence and dynamical exponent having unique values \cite{annni}.
 
 These above observations indicate that the two dimensional Ising model with the competing interactions in both the
 vertical and horizontal directions may show rich dynamical behaviour. In the present work we have studied the dynamics
 of two dimensional BNNNI (Bi-Axial Next Nearest Neighbour Ising) model where in addition to the ferromagnetic 
 nearest neighbour interaction we have anti-ferromagnetic next nearest neighbour interaction in both $x$ and $y$ directions. 
 
 The paper is organised as follows : In section II, we have described the model and its known properties. In section
 III, we have given a list of the quantities we have computed for studying the dynamical evaluation. In section IV, we
 have discussed the dynamic behaviour in detail. The discussions and conclusions are 
 made in the last section. 

\section{THE MODEL}

\begin{figure}[h]
\vspace{1.8cm}
 \includegraphics[width= 2.2cm, angle = 0]{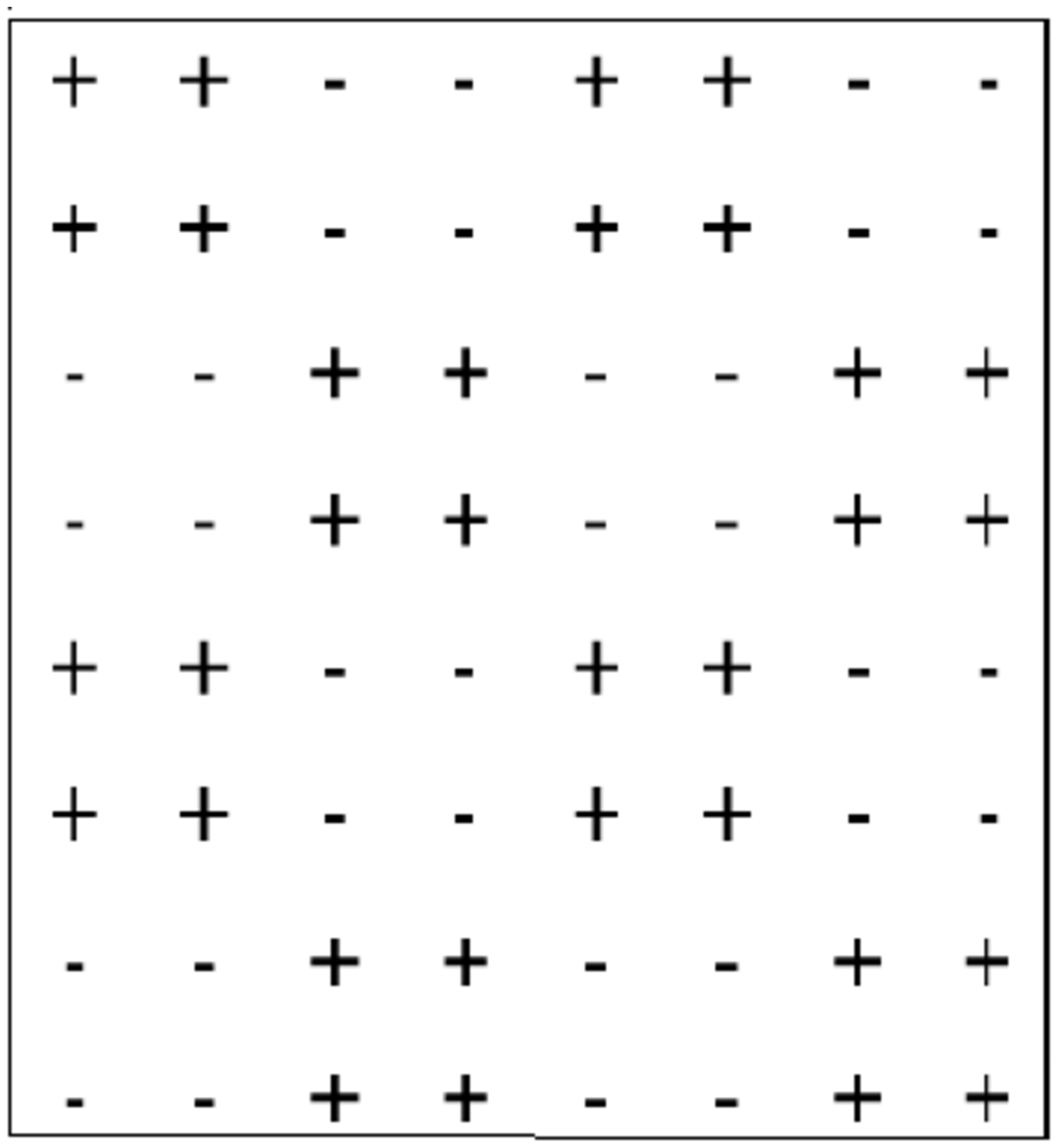}
 \hspace{1.3cm}
  \includegraphics[width= 2.2cm, angle = 0]{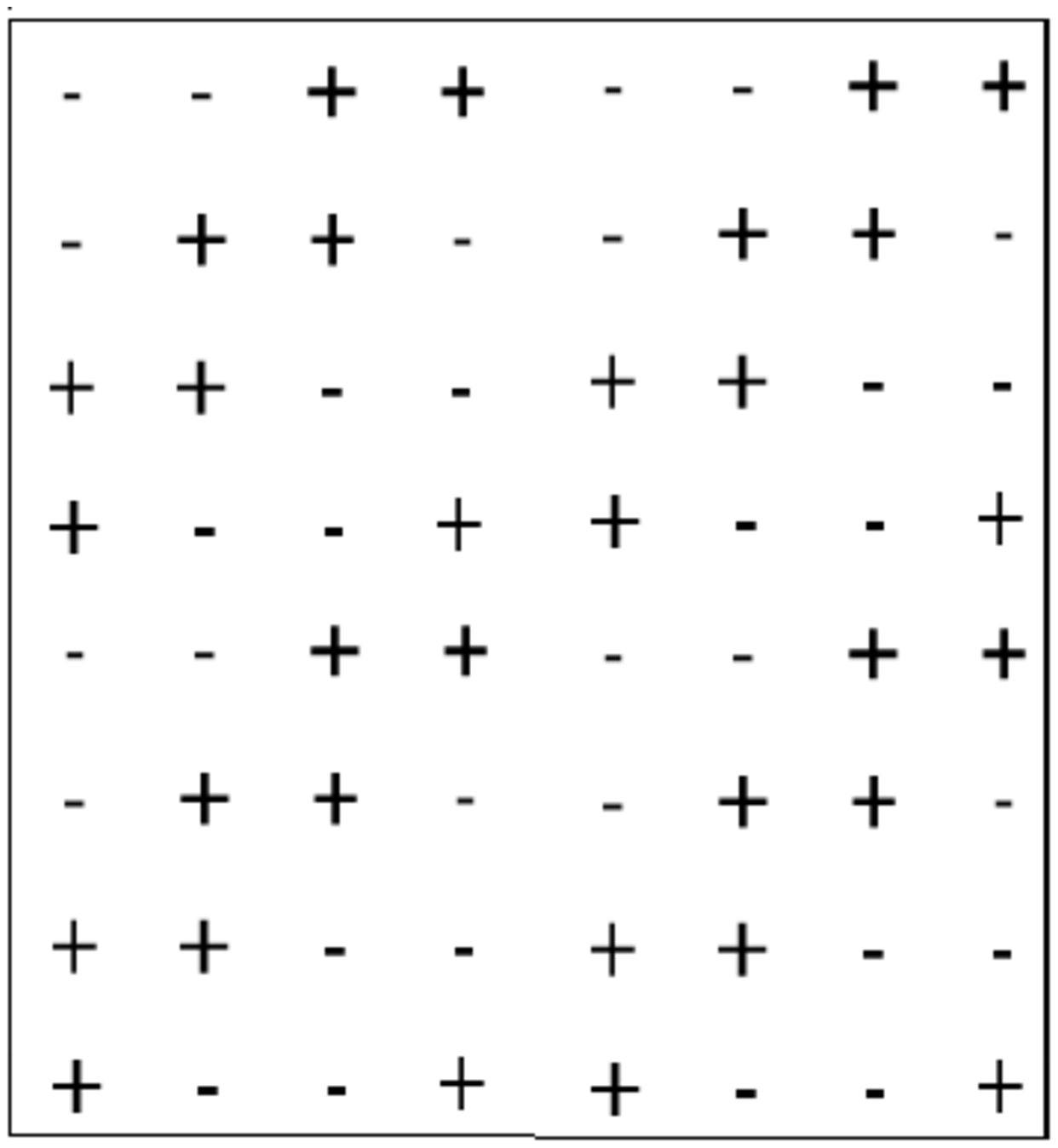}
  \caption{The antiphase ground state (temperature $T=0$) structures for $\kappa > 0.5$. First picture shows the `chessboard' type and the second one is `staircase' type.
  $+$ and $-$ signs stand for the up and down spin respectively.}
\label{grndst}
\end{figure}

The most generalised Hamiltonian for the two dimensional BNNNI model could be given by 
\begin{equation}
 H=-J_0\sum_{i,j=1}^L [(S_{i,j}S_{i+1,j}+S_{i,j}S_{i,j+1}) -\kappa (S_{i,j}S_{i+2,j} + S_{i,j}S_{i,j+2})]
 \label{bnnniH}
\end{equation}
where $\kappa$ is the ratio of the next nearest neighbour antiferromagnetic interaction and the nearest neighbour ferromagnetic interaction, which is same for both the $x$ and $y$ 
directions. The thermal phase diagram for two dimensional BNNNI model is not exactly known, but the ground state structures are known and quiet interesting. The ground state is 
completely ferromagnetic for $\kappa < 0.5$ and there exist antiphase order in both the vertical and horizontal direction for  $\kappa > 0.5$. However the ground state for 
$\kappa > 0.5$ can have two possible structures which contains the antiphase order in both the directions. These structures which have been illustrated in figure \ref{grndst}
are known as 'chessboard' (Fig \ref{grndst}(a) and 'staircase' (Fig \ref{grndst}(b)) configurations \cite{bnnni}.

Again, the ground state is infinitely degenerate at the fully frustrated point $\kappa = 0.5$

\section{QUANTITIES Computed}

To study and analyze the dynamical properties of the system following a zero temperature quench, we have computed the following quantities in this present work :
\begin{enumerate}
 \item Residual Energy $\varepsilon(t)$ : For studying the dynamics of this model the measurement of the domain walls are not very significant. 
 As the ground state have the antiphase order in both the directions (off course for $\kappa > 0.5$), the number of domain walls of the final state 
 are same as that of the average number of domain walls of the initial state and the number of domain walls do not decay with time. Though the number of domain walls remains almost constant, the energy of the system changes with time (for any value of $\kappa$) and the system self organizes itself to find out its minimum energy state. Hence instead of the number of domain walls, 
 the appropriate measure for studying this ordering dynamics is the residual energy per spin $\varepsilon = E - E_0 $, where $E_0= - 4J_0(1-\kappa)$ (for $\kappa < 0.5$) and 
 $E_0 = - 4J_0\kappa$ (for $\kappa > 0.5$) are the known ground state energy per spin and $E$ is the energy of the dynamically evolving state. 

The presence of domain walls in regular lattices causes an energy cost \cite{boyer}. It have been shown before  for the two dimensional Ising model that the residual energy 
have same scaling as that of the number or fraction of domain walls \cite{resising, bsnet}
\[
 \varepsilon \sim t^{-1/z}
\]
Hence computing the residual energy one can determine the value of the dynamical exponent $z$. Here we will call $z$ as ordering exponent, in stead of the domain growth exponent. 
\item Persistence probability $P(t)$: As mentioned earlier in the introduction, it is the probability that a spin does not flip until time $t$. If persistence probability decays in
a power law with time, that is $P(t) \sim t^{-\theta}$, the scaling form which can be used for finite size scaling is as follows \cite{puru, bsmodel}: 
\begin{equation}
P(t,L) \sim t^{- \theta}f(L/t^{1/z}).
\label{fss}
\end{equation}
For $x <<1$ the scaling function $f(x)  \sim x^{-\alpha}$ with $\alpha = z\theta$.  For large $x$, $f(x)$ is a constant.
Hence it is clear that for finite systems, the persistence probability saturates and the saturation value $P_{sat} \sim L^{-\alpha}$ at large times ($t \rightarrow \infty $). 
The value of the exponent $z$ obtained from the scaling of the residual energy should satisfy the scaling relation given by equation \ref{fss}. 

It has been previously shown that the exponent $\alpha $ is related to the fractal dimension of the
fractal formed by the persistent spins \cite{puru}. The fractal dimension $d_f= d -\alpha$, where $d$ is the dimension of the system. Here we have obtained an estimate 
of $\alpha$ and hence $d_f$ using the above scaling form of equation \ref{fss}.
\item Autocorrelation $A(t)$ : The autocorrelation function measures the correlation of the state of a single spin at time $t$ with its state at a previous time. 
The functional form of it is defined as : 
 \begin{equation}
  A(t)=\frac{\langle S_{i}(t)S_{i}(t_0)\rangle _{i}- \langle S_{i}(t)\rangle _{i} \langle S_{i}(t_0)\rangle _{i}}{
  \sigma_{i} (t) \sigma_{i} (t_0)}
  \label{auto}
 \end{equation}
 Where $S_{i}(t)$ and $S_{i}(t_0)$ is the state of the spin $i$ at time $t$ and $t_0$ respectively. $\langle \rangle _{i}$ is the average over $i$ index; and 
 $\sigma_{i} ( )$ is the standard deviation over $i$ index. We have studied the decay of autocorrelation for the system with the initial time only. That is $t_{0}=0$ 
for the equation \ref{auto}. 

For the nearest neighbour Ising model the auto-correlation function scales as \cite{autocorr}:
\begin{equation}
    A(t)\sim t^{-\lambda/z}
    \label{auto_expo}
\end{equation}
where $\lambda$ is the autocorrelation exponent and $z$ is the ordering exponent same as that of given by the scaling of residual energy and equation \ref{fss}. 
We have studied the decay of the autocorrelation function not only for the BNNNI model, but also for ANNNI spin system (Equation \ref{annni2d}) for different values of $\kappa$. 
\item Probability that the system will \textit{not} reach the ground state $P_{str}$ and the probability that the system will remain in the active state 
after very long time $P_{act}$ : These quantities have been computed by computing the percentage of the configurations which have not reached the 
ground state starting from a initial random state $P_{str}$ and the percentage of the configurations which remained active after a very long time $P_{act}$. 
\end{enumerate}

We have taken  lattices of size $ L\times L$ with $L=40$,$~80$,$~132$ and $200$ to study the dynamical behaviour of the system. 
The behaviour of different quantities which decays with time (residual energy, persistence and autocorrelation) have been averaged over 
at least 500 configurations for each lattice sizes. For estimating $P_{str}$ and $P_{act}$, we have averaged over much larger 
number of initial configurations (of the order of 4000). Periodic boundary condition has been used in both $x$ and $y$ directions.
 $ J_{0}= 1$ has been used in the numerical simulations.

\section{ DYNAMICAL BEHAVIOUR in detail} 

Before discussing the dynamical behaviour in detail, let us first discuss the stability of simple spin configurations which will help us to realise that the dynamical behaviour of the system is strongly dependent on $\kappa$.

\subsection{Stability of simple spin structures}
It is more or less well understood that the zero temperature dynamics of Ising spin system are mostly determined by the stability of spin configurations locally. Hence the system can
freeze with such spin configurations which do not correspond to global minimum of energy, but still can be stable dynamically. The well known example of this is the presence of striped state for two or higher dimensional Ising model with nearest neighbour interaction and ANNNI model (for $\kappa > 1$, equation \ref{annni2d}). Though these configurations are stable but do not correspond to the global minimum of energy. 

For the dynamics of the BNNNI model, the fate of a randomly selected spin is determined by the state of its eight neighbours which could be in any one of the $2^8=256$ possible configurations. 
Though for many of these configurations the dynamical behaviour of the central spin is similar for any given value of $\kappa$. For example, if among the four nearest neighbours, two of them are up and rest two are down or vice versa, the state of central spin will be determined by the configurations (which could be any one of the $2^4 \times 6=96$ configurations) of the next nearest 
neighbours. For the similar configurations of the next nearest neighbours, the dynamics will be determined by the orientations of the nearest neighbours only. Except these cases for most of the configurations, the fate of a randomly selected spin will depend on the values of $\kappa$ and we will see that one can expect the similar dynamical behaviour for a range of $\kappa$ values. We should also remember that for having a stable configuration locally, not only the randomly selected spin, but also its eight neighbours have to be stable. 

\begin{figure}[ht]
 \includegraphics[width= 6.6cm, angle = 0]{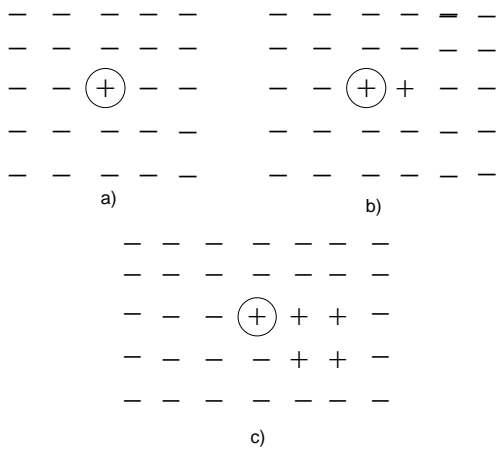}
  \caption{Schematic diagram of the spin configurations for the analysis of stability of simple structures : The conditions for the stability of the central spin (which have been circled) and its neighbours have been discussed in the text.}
\label{schematic}
\end{figure}
First let us consider the simplest configuration of a single up spin in a sea of down spins (Fig. \ref{schematic}(a)) or vice versa. The central spin will be stable for $\kappa > 1$ and is unstable for $\kappa < 1$. For  $\kappa = 1$ the spin will flip with probability $1/2$, hence making the dynamics stochastic. However all its four nearest neighbours are stable only for 
$\kappa < 0.5$. On the other hand four next nearest neighbours of the central spin are stable when $\kappa < 2$. 
Now for a domain of two up spin in a sea of down spins (as shown in Fig\ref{schematic}(b) ), both the up spins will be stable for 
 $\kappa > 0.5$. But the down spins at their nearest neighbours (like the three nearest neighbours of the circled spin which are at the down state) will be stable when $\kappa < 0.5$. On the other hand for the four next nearest neighbours, three of them will be stable as long as $\kappa < 2$ and one of then will be stable for $\kappa < 1$.

Next we would like to consider the configuration of a domain of five up spins in a sea of down spins (Fig.\ref{schematic}(c)). The spin at the corner of the domain of up spins (it have been circled), will be stable for $\kappa > 1$ only. The up spin at the right
nearest neighbour of this circled spin will be always stable for any value of $\kappa > 0$. But the down spin at the left nearest
neighbour will be stable only for $\kappa < 1$. The other two down spin at the nearest neighbour of the circled spin will be stable
as long as $\kappa < 0.5$. Among the next nearest neighbours of the circled spin, the up spin will be stable for any value of 
$\kappa > 0$, but all the down spins will be stable whenever $\kappa < 2$. 

One can consider more complicated structures but the analysis of these simple structures indicates that there could be different dynamical behaviour in the region $\kappa < 1$, $\kappa = 1$, $\kappa > 1$, $\kappa =2$ and $\kappa >2$. However as far as the dynamical behaviour of the system 
(that means the behaviour of all the quantities we have computed including residual energy,  persistence autocorrelation function etc) is concerned, we find that there exists only three regions with different dynamical behavior : $\kappa < 1$, $\kappa = 1$ and $\kappa > 1$. 

We have also studied the decay of the autocorrelation function for the ANNNI model (Hamiltonian is given by equation 
\ref{annni2d}), which have been discussed and compared with BNNNI at the last subsection of this section. 

\subsection{ Dynamics in the region $0 < \kappa < 1$ }
Though from the analysis of the simple spin structures, one can guess that the dynamical behaviour could be different for $\kappa < 0.5$ and $\kappa > 0.5$, but we find that the system has identical dynamical behaviour for all $\kappa$, in the region $0 < \kappa < 1$. 
At this parameter regime, the system does not go to its equilibrium ground state at all making $P_{str}$ to be $one$ for all the values of $\kappa$. At the beginning of the dynamics, domains of size one will 
vanish rapidly. After that for the above mentioned reasons, the dynamics will be bit complex and slow, but will continue for a long time. 

\begin{figure}[ht]
 \includegraphics[width= 7.5cm, angle = 0]{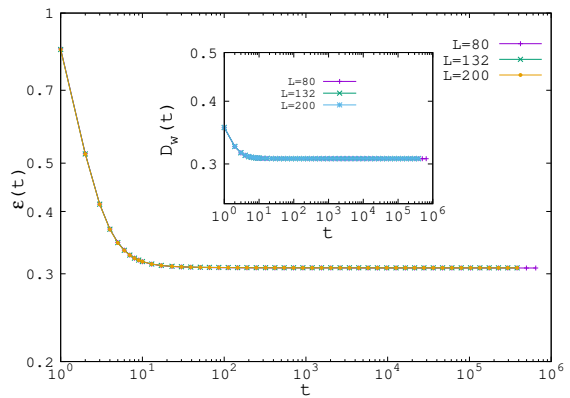}
  \caption{Decay of the residual energy $\varepsilon(t)$ with time for $\kappa=0.6$. Fraction of domain walls as a function of time for the same value of $\kappa$ have been plotted in the inset.}
\label{dom_res}
\end{figure}
 Fraction of domain walls decay very little in both the directions at the initial time and then remain constant for the rest of the dynamics (Inset of Fig. \ref{dom_res}). The residual energy also stop decaying after some time but the dynamics continues.
 It is prominent from figure \ref{dom_res} that finite size effect does not exist for the decay of residual energy and the fraction of domain walls. 

\begin{figure}[ht]
 \includegraphics[width= 7.2cm, angle = 0]{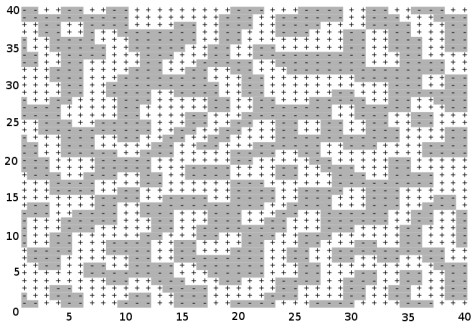}
  \caption{Snap shot of $40 \times 40 $ lattice at time $t=10$ for $0 < \kappa < 1$. 
  $+$ and $-$ signs stand for the up and down spin respectively.}
\label{snap1}
\end{figure}

\begin{figure}[ht]
 \includegraphics[width= 6.9cm, angle = 0]{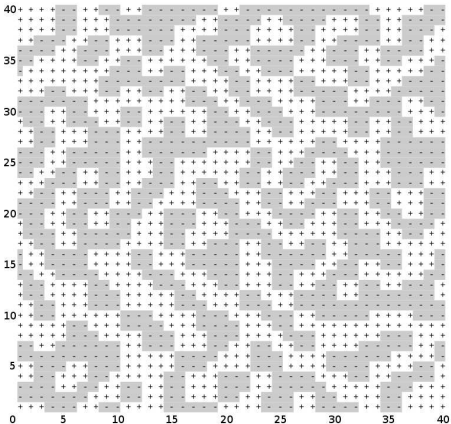}
  \caption{Snap shot of $40 \times 40 $ lattice at time $t=50$ for $0 < \kappa < 1$. 
  $+$ and $-$ signs stand for the up and down spin respectively.}
\label{snap2}
\end{figure}

\begin{figure}[ht]
 \includegraphics[width= 6.9cm, angle = 0]{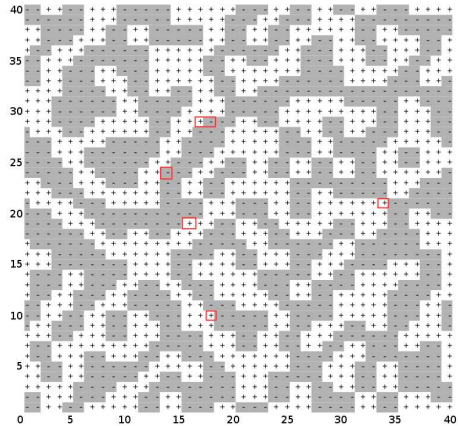}
  \caption{Snap shot of $40 \times 40 $ lattice at time $t=500$ for $0 < \kappa < 1$. 
  $+$ and $-$ signs stand for the up and down spin respectively.}
\label{snap3}
\end{figure}

\begin{figure}[ht]
 \includegraphics[width= 6.9cm, angle = 0]{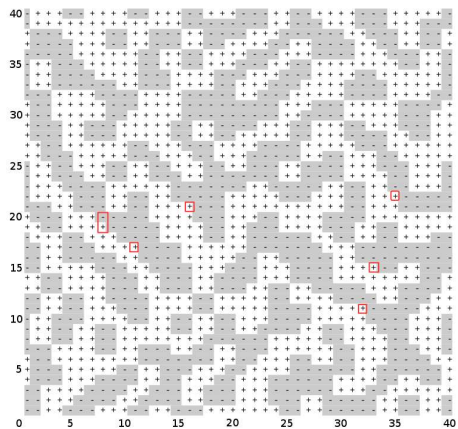}
  \caption{Snap shot of $40 \times 40 $ lattice after a very long time that is at $t \rightarrow\infty (at $ $t=500000$) for $0 < \kappa < 1$. $+$ and $-$ signs stand for the up and down spin respectively.}
\label{snap4}
\end{figure}

 Figs. \ref{snap1}, \ref{snap2}, \ref{snap3} and \ref{snap4} are the snapshots of the system 
at different times which shows the evolution of the typical lattice structures at this parameter regime after a quench to zero temperature.

From very early time, diagonal strips will appear in the lattice which will remain forever. These diagonal strips become more and more prominent in the lattice with time as the dynamics continues even after the residual energy stop decaying.

After some initial time (for $t>300$ Fig. \ref{dom_res} ), spin flips do not reduce the energy of the system and hence the 
residual energy stop decaying. Though the energy of the system does not decay anymore, few of these spins remain active for very long time, even at $t\rightarrow \infty$ making $P_{act}=1$ for any values of $\kappa <1$. Some of those spins have been high lightened in the red colour box in figure \ref{snap3} and \ref{snap4}.

The active sites (the sites at which the spins keep flipping) move throughout the lattice along the edge of the diagonal strips, killing the persistence of all the sites of the lattice.  
Persistence probability for $\kappa <1$ shows a slow decay with time and goes to $zero$ at long time.
\begin{figure}[ht]
 \includegraphics[width= 5.2cm, angle = 270]{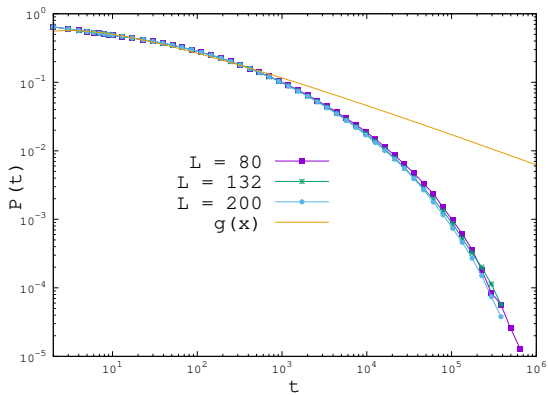}
  \caption{Decay of the persistence probability with time for $\kappa<1$. The fitting of the approximated functional form (the form of the function $g(x)$ is given in the text) have been shown at the beginning of the dynamics for an appreciable range of time.}
\label{perklt1}
\end{figure}
 The functional form for the decay is different at the beginning and at the end of the dynamics. At the beginning of the dynamics the decay is slower than that of the later time and the functional form can be approximated as $g(x) \sim t^{-c} \times ln(bt)$ for an appreciable range of time. We have numerically found that the function $g(x)$ fits well at the beginning of the dynamics with $b \simeq 2.266$ and $c \simeq 0.515$ (Figure. \ref{perklt1}). 
However at late time, it is not possible to characterise the decay of the persistence probability by some simple mathematical function of time. The decay of the persistence probability do not have any effect of the finite system sizes (Figure. \ref{perklt1}).

We have also studied the decay of the auto-correlation function with time for this parameter region. The results for that have been presented at the last subsection of this section. 

\subsection{ Dynamics for $\kappa > 1$ }
\label{dynkgt1}

It had been previously observed for axially next nearest neighbour Ising model with the competing interactions (that is ANNNI model) that the dynamical behaviour changes drastically at $\kappa =1$ in both one and two dimensions.
For two dimensional BNNNI model, the presence of competing interaction in both the directions have the effect to the dynamics in a large extent. As mentioned before, like ANNNI model the dynamical behaviour of BNNNI model is also different
for  $\kappa = 1$ and  $\kappa > 1$. In this section we will discuss the dynamical behaviour of the mode for  $\kappa > 1$. 

\begin{figure}[h]
 \includegraphics[width= 7.8cm, angle = 0]{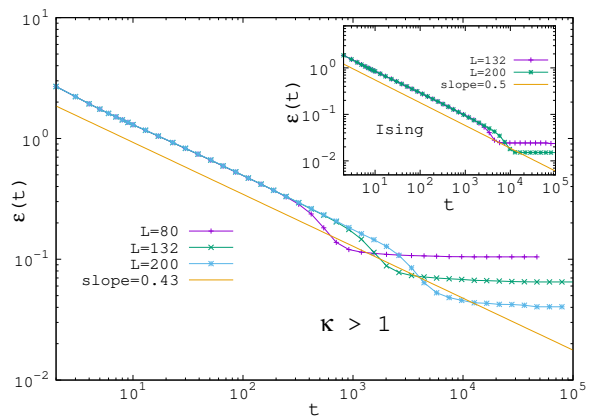}
  \caption{Decay of the residual energy with time for $\kappa>1$. The amber color line have slope 0.43 in the main plot. 
  Inset shows the decay of the residual energy with time for two dimensional nearest neighbour Ising model.The amber color line have slope 0.5 at the inset.}
\label{reskgt1}
\end{figure}

The number of domain walls do not much change with time and quickly saturates at an value $0.5$ for the fraction of the domain walls, in both the horizontal and vertical direction. 
Though the domain walls saturates very early, residual energy decay in a power law in time, with a decay exponent $\sim 0.43$ (Figure \ref{reskgt1}). 
This yield the ordering exponent $z\simeq 2.33$ as  $\varepsilon \sim t^{-1/z}$. In the inset of figure \ref{reskgt1}, we have also shown the decay of the residual energy for the two dimensional nearest neighbour Ising model.

 It is clear from the saturation of the residual energy that some kind of stripped states exist in the system in this parameter
 regime. Just before the saturation, residual energy shows an exponential decay for some small time. This is due to the
 exponential decay of $\varepsilon$ for those configurations which go to the ground state in stead of getting freezed in one of the stripped states. Saturation of residual energy (and an exponential decay of it just before the saturation) for two dimensional nearest neighbour Ising model [inset of figure \ref{reskgt1}] is also due to the presence of the stripped state in the lattice \cite{Krap_Redner}. 
 
 Persistence probability decays in a power law in time and the persistence exponent $\theta \simeq 0.22$. 
 One can expect $z\simeq 2.33$ from the finite size scaling analysis, if the ordering exponent $z$ is similar to that of the previously known domain growth exponent. 
 \begin{figure}[h]
 \includegraphics[width= 7.9cm, angle = 0]{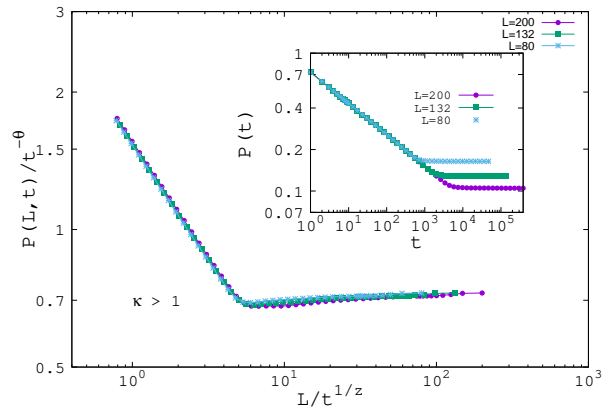}
  \caption{The collapse of scaled persistence data versus scaled time using $\theta = 0.22$ and  $z=2.33$ is shown for different
system sizes for $\kappa > 1$. Inset shows the unscalled data for the decay of the persistence probability with time. }
\label{perkgt1}
\end{figure}
 We indeed found $\theta = 0.22 \pm 0.002$ and $z=2.33 \pm 0.01$ performing the finite size scaling analysis following equation (\ref{fss}) and we have checked this for different values of $\kappa$ ($\kappa = 1.3, ~ 1.6, ~2.0, ~2.5 $ and $20$). 
 So we conclude that these exponents are independent of $\kappa$ for $\kappa > 1$ and also the ordering exponent $z$ is similar to the dynamical exponent which is previously known as the domain growth exponent. A typical behaviour of the
raw persistence data as well as the data collapse for the finite size scaling is shown in Fig. \ref{perkgt1}.

Next we asked what is the probability $P_{str}$, that the system will \textit{not} reach the ground state and will freeze in one of the stripped state. This have been calculated by computing the fraction of the initial configurations 
which couldn't reach the ground state (those states have $non-zero$ residual energy) at all after a very long time.  

 \begin{figure}[h]
 \includegraphics[width= 5.5cm, angle = 270]{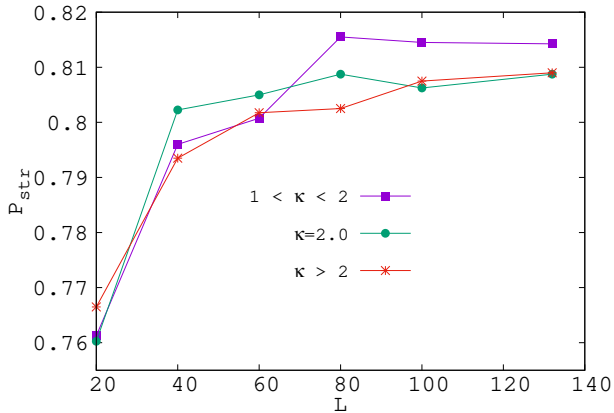}
  \caption{Freezing probability $P_{str}$ is plotted with the system size for different range of $\kappa$ values. }
\label{fss_grndkgt1}
\end{figure}
We found that little more than 80$\%$ configurations freeze to some stripped states before reaching the ground state. 
The freezing probability $P_{str}$ initially increases with system sizes and then saturates for larger systems [Fig \ref{fss_grndkgt1}]. We didn't found any significant differences in the values of  $P_{str}$ for larger sizes for $\kappa > 1$. 
It may appear from figure \ref{fss_grndkgt1}, that the freezing probability have different saturation values (the values for 
large enough system sizes) for $ 1 < \kappa < 2$ and $\kappa \geq 2$, but the difference between these two values are 
insignificant ($P_{str} \simeq 81.5\%$ when $ 1 < \kappa < 2$ and $P_{str} \simeq 81\%$ for  $\kappa \geq 2$). We also found that the configurations which reach the ground state, half of them reach the checker board configuration and the other half reach the staircase configuration. 

$P_{act}$ the probability of being in the active state after a very long time is equal to $zero$ for any values of $\kappa$ for $\kappa > 1$. That means all the configurations either go to the stripped state (with no active state or site in it) or to one of the ground states at $t\rightarrow\infty$.

\begin{figure}[h]
 \includegraphics[width= 7.4 cm, angle = 0]{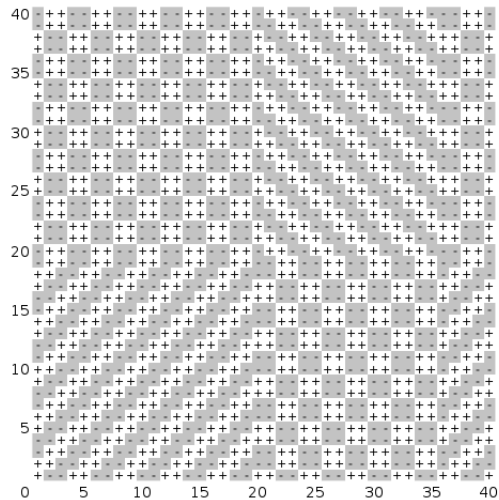}
  \caption{A typical snapshot of the stripped state for $\kappa > 1$ after the system reaches the steady state configuration.}
\label{strpkgt1}
\end{figure}
It is not straight forward to imagine the structure of the stripped states appear in the lattice after it reaches the steady state at the end of the dynamics. We found that the checker board configuration and the staircase configuration stay together in the lattice. The energy cost at the interface of these two configurations is more than the energy of the ground state, though these interfaces are stable for $\kappa > 1$. A typical snap-shot of this type of stripped state which appear for $\kappa > 1$ have been shown in figure \ref{strpkgt1}. 

Also for this parameter region, we have studied the decay of the auto-correlation function with time. However the results for that have been presented at the last subsection of this section.  

\subsection{ Dynamics for $\kappa = 1$ }
\label{dynk1}

\begin{figure}[ht]
 \includegraphics[width= 5.5 cm, angle = 270]{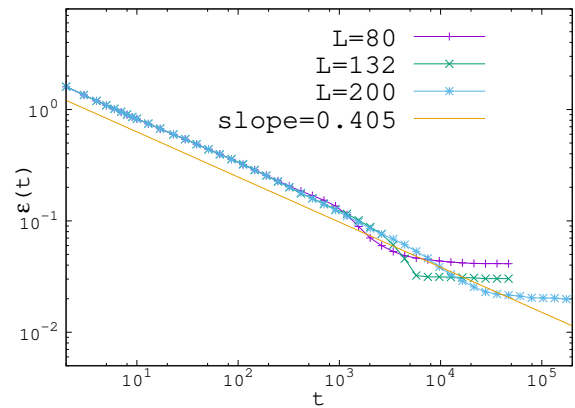}
  \caption{Decay of the residual energy with time for $\kappa=1$. The amber color line have slope 0.405 .}
\label{reskeq1}
\end{figure}

Residual energy decays in a power law with time and from the decay we found that the ordering exponent $z$ to be close to $2.47$ [Figure \ref{reskeq1}] which is significantly different from the value of $z$ for $\kappa>1$. 
Saturation value of the residual energy indicates that also for $\kappa = 1$, the stripped state exists in the system after it reaches the steady state. This phenomena is contrary to what happens in 2-d ANNNI model \cite{annni} for $\kappa = 1$ where the system reaches the ground state with probability $one$. 

Persistence probability decays in a power law with the exponent $\theta = 0.332 \pm 0.002$. The finite size scaling  
\begin{figure}[ht]
 \includegraphics[width= 7.6 cm, angle = 0]{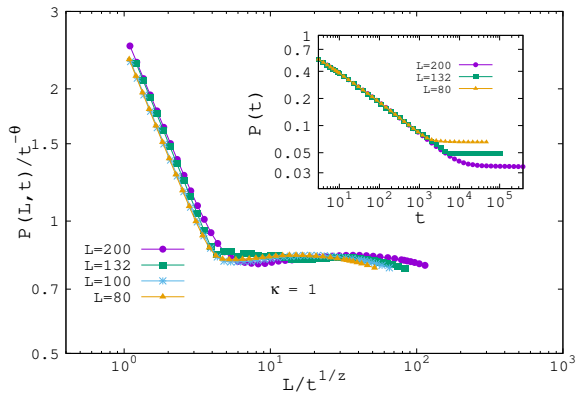}
  \caption{The collapse of scaled persistence data versus scaled time using $\theta = 0.332$ and  $z=2.48$ is shown for different system sizes for $\kappa = 1$. Inset shows the unscalled data for the decay of the persistence probability with time. }
\label{perkeq1}
\end{figure} 
suggests the ordering exponent $z=2.47 \pm 0.04$ and the value is consistent with what we get from the decay of residual energy. A typical behaviour of the raw persistence data as well as the data collapse using the finite size scaling analysis 
is shown in Fig. \ref{perkeq1}. The quality of the data collapse is $not$ as good as that of $\kappa > 1$, which gives 
a higher error bar on the ordering exponent $z$.

We found that less than $1\%$ of the configurations remain active after a very long time for the higher sizes making $P_{act}$ to be very low. The probability for a configuration to be in an active state is almost zero for the lower system sizes. Just like $\kappa > 1$, here also we asked about the probability  ($P_{str}$) that a system will not reach the ground state. 
\begin{figure}[ht]
 \includegraphics[width= 7.6 cm, angle = 0]{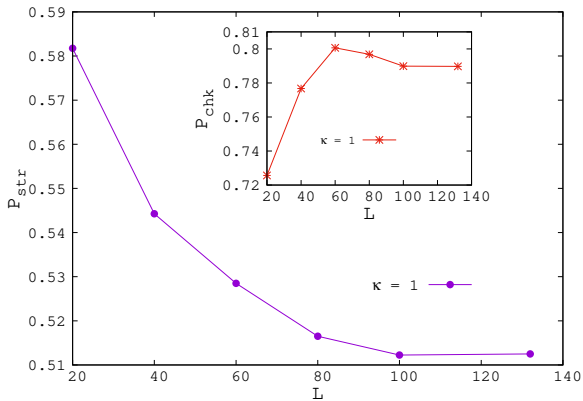}
  \caption{Freezing probability $P_{str}$ is plotted with the system size for  $\kappa = 1$. Variation of $P_{chk}$, the
  probability for a system to reach the checker board configuration if it reaches the ground state, have been plotted against the system size $L$ at the inset.}
\label{fss_grndstkeq1}
\end{figure}
In this case either it will freeze in one of the stripped states or will end up in one of the rare active states. 
We found that little more than $50\%$ configurations fail to reach the ground state even after a very long time [Figure \ref{fss_grndstkeq1}]. The configurations which reach the ground state, surprisingly almost $80\%$  (being precise $79\%$ according to the numerical estimate we obtained) of them reach the checker board configuration and little more $20\%$ 
configurations reach the staircase configuration [Inset of Figure \ref{fss_grndkgt1}]. 
 
As we have mentioned earlier, in case of large system sizes very few configurations can remain active even after very long time. This long lived configurations are actually diagonal stripe configuration which had been previously observed to appear 
in two dimensional Ising model \cite{Krap_Redner}. But the dynamics for this configurations are much more complicated than that of the two dimensional Ising model. This configurations will eventually go to the ground state. But it is not simple to argue, how much time is required for this configurations to reach the ground state.  

 \begin{figure}[ht]
 \includegraphics[width= 7.5 cm, angle = 0]{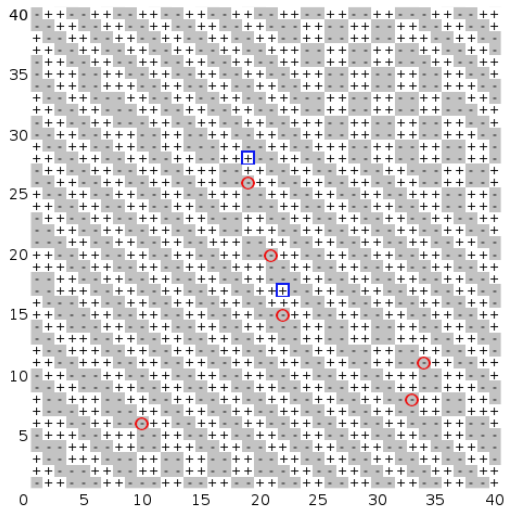}
  \caption{A typical snap shot of an active configuration after a very long time for $\kappa = 1$. Few of the active sites have been circled in red color.}
\label{actconf}
\end{figure}
 Though it is not trivial to detect these rare configurations, we have shown a typical snap shot of this kind of active configuration in figure \ref{actconf}. Usually all the neighbours of the active lattice sites (circled in red color in Fig. \ref{actconf}) are stable except one. If the active site flip (it will flip will probability $1/2$ as the energy for that site is $zero$), the unstable neighbour become stable and one stable neighbour become unstable (shown in blue square in Fig. \ref{actconf}). This is how the active sites move in the lattice in a pair unless a local configuration for the definite flip have formed in the lattice. The mechanism make the dynamics very slow at this point of time. Time taken by the system to reach
 the ground from this kind of active configuration is order of magnitude higher than other configurations and computing the time is beyond the available computational power. 
 
 \subsection{Decay of auto-correlation function}
 In this section we present the results for the decay of autocorrelation function with time for different values of 
 $\kappa$. To compare the results we have studied the time decay of the function also for ANNNI model (given by equation
 \ref{annni2d}), as that have not been studied before in \cite{annni}. For two dimensional nearest 
 neighbour Ising model (which corresponds to $\kappa=0$), the value of the autocorrelation exponent $\lambda \simeq 1.25$ \cite{autocorrising} have been verified from our numerics. 
 
 \begin{figure}[ht]
 \includegraphics[width= 7.5 cm, angle = 0]{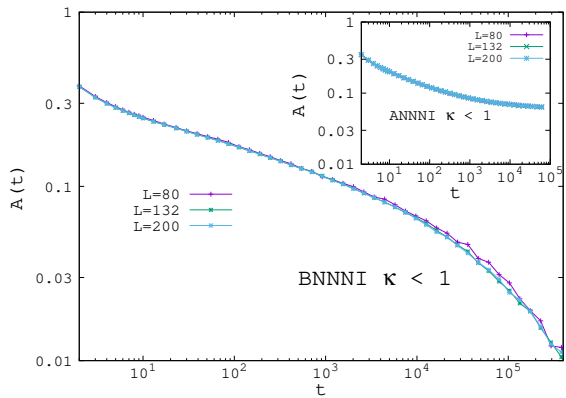}
  \caption{Decay of autocorrelation function with time for $\kappa < 1$. Inset shows the same for the ANNNI model. }
\label{autocorr_klt1}
\end{figure}

 For $\kappa < 1$, the autocorrelation function does not decay in a power law fashion for both the BNNNI and ANNNI model (Figure. \ref{autocorr_klt1}). 
 For BNNNI model, autocorrelation function slowly decays to zero; though it is almost impossible to write any simple mathematical form for the decay of the function with time. 
 For ANNNI model autocorrelation function also shows a slow decay with time, however that can be approximated by $1/log(t)$ for an appreciable range of time. At larger time, it saturates at a finite value unlike BNNNI model. 
   
  \begin{figure}[ht]
 \includegraphics[width= 7.5 cm, angle = 0]{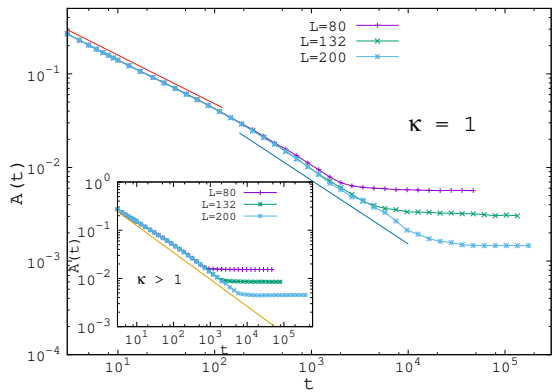}
  \caption{Decay of the autocorrelation function with time have been plotted for $\kappa=1$ in the main plot. The slope of the red line at the beginning is $0.52$ and the slope of the blue line in the same plot is $0.68$. Inset shows the decay for $\kappa > 1$, where the slope of the yellow line is $0.558$.}
\label{autocorr_BNNNI_kge1}
\end{figure}
 For $\kappa \geq 1$ autocorrelation function shows a power law decay with time, with different the decay exponent for 
 $\kappa =1$ and $\kappa > 1$. For $\kappa > 1$, it decays as $ t^{-\eta}$ with $\eta = 0.558 \pm 0.002$ 
after some initial time [inset of Fig. \ref{autocorr_BNNNI_kge1}]. This yields the autocorrelation exponent $\lambda =1.3 \pm 0.01$ as  $\eta =\lambda/z$ [Eq. \ref{auto_expo}] as $z=2.33 \pm 0.01$ [section \ref{dynkgt1}]. However for 
$\kappa=1$, the decay exponent appears to be different for some initial time and at the large time. At the beginning it
decays as $ t^{-\eta}$ with $\eta = 0.521 \pm 0.002$ giving $\lambda =1.29 \pm 0.025$ as $z=2.47 \pm 0.04$ [section \ref{dynk1}]. At long time the decay exponent $\eta = 0.68 \pm 0.002$ which corresponds the autocorrelation exponent $\lambda =1.68 \pm 0.03$.

On the other hand in case of ANNNI model for $\kappa \geq 1$ autocorrelation function have a power law decay with same decay exponent $\eta$ for both  $\kappa =1$ and $\kappa > 1$. 
  \begin{figure}[ht]
 \includegraphics[width= 7.5 cm, angle = 0]{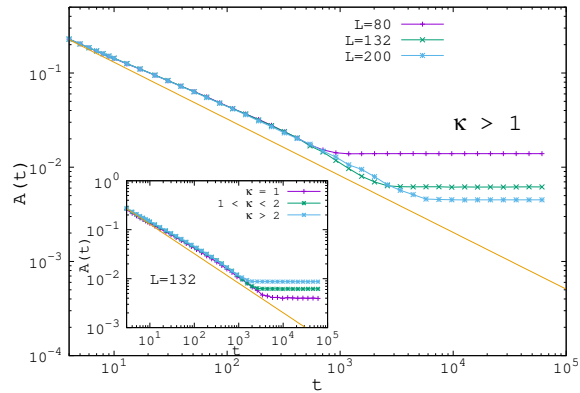}
  \caption{Decay of the autocorrelation function with time for different sizes have been plotted for $\kappa > 1$ in case of ANNNI model. Inset shows that the decay exponent is same 
  for any value of $\kappa \geq 1$ for ANNNI. The slope of the amber colored line is $0.602$ for both the main plot and inset. }
\label{autocorr_ANNNI_kge1}
\end{figure}
However as the value of $z$ is different for $\kappa =1$ and $\kappa > 1$ \cite{annni}, the autocorrelation exponent will be different. We found $ A(t) \sim t^{-\eta}$ with $\eta = 0.602 \pm 0.002$ for $\kappa \geq 1$. That concludes 
$\lambda = 1.11 \pm 0.01 $ for $\kappa = 1$ (as $z=1.84 \pm 0.01$ \cite{annni}) and  $\lambda = 1.25 \pm 0.01$ 
for $\kappa > 1$ (as $z=2.08 \pm 0.01$ \cite{annni}) for the ANNNI model. 

\section{Discussions and Conclusions}
We have studied the dynamical features of BNNNI model in two dimensions following a quench to zero temperature.
We have seen that the dynamics is very much dependent on the value of $\kappa$, the ratio of the antiferromagnetic to the ferromagnetic interaction in both the directions. Depending on the dynamical features we can distinguish three different
regime $\kappa < 1$, $\kappa =1$ and $\kappa > 1$ just like ANNNI model in two dimension. Though the intrinsic dynamical
behaviour is of the model is bit different from that of ANNNI. Presence of the competing interaction in both the vertical
and horizontal directions (which make the model symmetric unlike ANNNI) can affect the dynamics substantially. 
For example unlike ANNNI model, here the system remain in the active state forever when $\kappa <1$. 

For studying the dynamics of ordering after a quench to zero temperature, we have studied the decay of the residual energy, as domain walls do not decay with time. When residual 
energy have power law decay (that is for $\kappa \geq 1$), we claim that the decay exponent is similar to that of the domain growth exponent $z$. He have obtained the data collapse 
of persistence data for $\kappa \geq 1$ successfully, using the value of the dynamical exponent $z$ acquired from the power law decay of residual energy. 
As the system organises itself to find out its minimum energy state despite of the fact that the traditional domain growth phenomena does not happen, we have called the exponent 
$z$ as ordering exponent. 

For $\kappa > 1$, we found the persistence exponent $\theta$ to be same as that of the two dimensional nearest neighbour
Ising model (corresponds to $\kappa=0$), though the value of the ordering exponent $z$ is bit different. This makes the exponent $\alpha=z\theta$ to be very 
different for $\kappa=0$ and $\kappa>1$. For $\kappa=0$, $\alpha \simeq 0.44$ while for $\kappa>1$, $\alpha = 0.512 \pm 0.005$. This tells us that the spatial 
correlations of the persistent spins are quite different for the two and not only the the dynamical class for $\kappa=0$ and $\kappa>1$ are different (which is 
already evident from the difference in the value of $z$), also the persistence behaviour is not the same. $\kappa=1$ appears to be a special point where the 
dynamic behaviour changes radically with a different value of $\theta$ and $z$ than that of $\kappa = 0$ and $\kappa > 1$. Here we got the value  of $\alpha$ to be 
$0.82 \pm 0.018$. The error bar is higher as an effect of having a higher error bar on $z$ for $\kappa=1$. 

Next we would like to comment on the behaviour of the autocorrelation function. We have studied the decay of the function with time from which we have obtained the values of the autocorrelation exponent $\lambda$ 
(for $\kappa \geq 1$, when the function have a power law decay), using the value of $z$. We have studied this not only
for BNNNI, but also for ANNNI model, which is anisotropic by nature. For BNNNI model, when $\kappa > 1$, the value of
$\lambda$ is close but different from that of $\kappa=0$. However for ANNNI, the value of $\lambda$ obtained from our
nemurics is similar to the two dimensional nearest neighbour Ising model. For $\kappa=1$, the autocorrelation exponent 
$\lambda$ appears to be different at the beginning and at the end of the dynamics. At the beginning $\lambda$ is almost
same as that of $\kappa > 1$, though at late time the value of $\lambda$ is very different and higher than that. 
On the other hand, in case of ANNNI model for $\kappa=1$, the value of $\lambda$ is lower than that of $\kappa > 1$ 
and $\kappa=0$.

It is important to note that the system always reach an absorbing state for $\kappa > 1$ and remain in the active state 
when $\kappa < 1$. For $\kappa=1$ most of the configurations go to the absorbing state expect a very few long lived
configurations. These rare configurations reach the absorbing state taking a very long time which is order of magnitude
higher than the time taken by other configurations. This indicates that there may exists an active to absorbing phase
transition around $\kappa =1$. This type of behaviour for zero temperature single spin flip dynamics have been observed
before for one dimensional ANNNI model (which is also isotropic by nature), where the system remain in the active state
for $\kappa < 1$ and reaches the absorbing state for $\kappa \geq 1$. Hence if an active to absorbing phase transition
exists, that can be checked and studied in detail as separate problem in future. 

The single spin flip dynamics for the BNNNI and ANNNI model can also be studied in three and higher dimension. The
dynamical structure of three dimensional nearest neighbour Ising model, after a quenching to zero temperature is already 
complex and bit interesting \cite{3dising}. Hence one can expect novel dynamical behaviour for three dimensional BNNNI 
and ANNNI model too. Also following the hypothesis which have been noted in the previous paragraph, there should not
exist any plausible active to absorbing phase transition in three dimension for these models as the both models do not 
remain isotropic in the higher dimension. 

In this paper, we have studied the dynamical behaviour of the two dimensional BNNNI model under a zero temperature 
quench. The dynamics at finite temperature can be quite different. As the spin flipping probabilities are stochastic at 
finite temperatures, and the dynamical frustration for which the system freezes before reaching its ground state, can be overcome by the thermal fluctuations. We would also like to note that, given the definition of persistence being bit 
different at finite temperatures \cite{derrida2}, it is not simple to guess the persistence behaviour (for any spatial 
dimension) just from the results of the zero temperature quenching dynamics. The single spin flip dynamics of the BNNNI
model after a quench to the finite temperature indeed remain as an open problem which could be addressed in future.
\medskip

Acknowledgments:
The authors thank Denis Boyer for useful discussions. Both the authors would like to acknowledge the financial support 
from SEP under Prodep program number 238628 folio UDG-PTC-1307. The computational power provided by CIC-UNAM for some 
of the simulations have also been acknowledged.

\end{document}